\documentclass[reprint,superscriptaddress,aps,prl]{revtex4-2}
\usepackage{color,graphicx,calc,url}
\usepackage{dcolumn}
\usepackage{bm}
\usepackage{amssymb}
\usepackage{amsmath}
\usepackage{wasysym}
\usepackage{mathrsfs} 
\usepackage{array}
\usepackage{mathtools}
\usepackage{xcolor}
\usepackage{commath}
\usepackage[colorlinks=true,
            linkcolor=red,
            urlcolor=blue,
            citecolor=blue]{hyperref}

\begin{document}
%\title{Quadratic and linear magnetooptic Kerr effect spectroscopy on partially ordered Co$_2$MnSi Heusler compounds}
\title{Cubic magneto-optic Kerr effect in Ni(111) thin films with and without twinning}

\affiliation{Center for Spinelectronic Materials and Devices, Department of Physics, Bielefeld University, 33615 Bielefeld, Germany}
\affiliation{IT4Innovations, V\v{S}B-Technical University of Ostrava, Ostrava 70800, Czech Republic}
\affiliation{Nanotechnology Centre, V\v{S}B-Technical University of Ostrava, Ostrava 70800, Czech Republic}
\affiliation{Faculty of Mathematics and Physics, Charles University, Prague 12116, Czech Republic}

\author{Maik Gaerner}
\thanks{contributed to this work equally}
\affiliation{Center for Spinelectronic Materials and Devices, Department of Physics, Bielefeld University, 33615 Bielefeld, Germany}

\author{Robin Silber}
\thanks{contributed to this work equally}
\affiliation{IT4Innovations, V\v{S}B-Technical University of Ostrava, Ostrava 70800, Czech Republic}
\affiliation{Nanotechnology Centre, V\v{S}B-Technical University of Ostrava, Ostrava 70800, Czech Republic}

\author{Tobias Peters}
\affiliation{Center for Spinelectronic Materials and Devices, Department of Physics, Bielefeld University, 33615 Bielefeld, Germany}

\author{Jaroslav Hamrle}
\affiliation{Faculty of Mathematics and Physics, Charles University, Prague 12116, Czech Republic}
\author{Timo Kuschel}
\email{tkuschel@physik.uni-bielefeld.de}
\affiliation{Center for Spinelectronic Materials and Devices, Department of Physics, Bielefeld University, 33615 Bielefeld, Germany}

%%\date{\today}


\begin{abstract}
\noindent In most studies utilizing the magneto-optic Kerr effect (MOKE), the detected change of polarized light upon reflection from a magnetized sample is supposed to be proportional to the magnetization~$\bm{M}$. However, MOKE signatures quadratic in $\bm{M}$ have also been identified and utilized, e.g., to sense the structural order in Heusler compounds, to detect spin-orbit torque or to image antiferromagnetic domains. In our study, we observe a strong anisotropic MOKE contribution of third order in~$\bm{M}$ in Ni(111) thin films, attributed to a cubic magneto-optic tensor $\propto $ $\bm{M}^3$. We further show that the angular dependence of cubic MOKE (CMOKE) is affected by the amount of structural domain twinning in the sample. Our detailed study on CMOKE will open up new opportunities for CMOKE applications with sensitivity to twinning properties of thin films, e.g. CMOKE spectroscopy and microscopy or time-resolved CMOKE.

\end{abstract}

\maketitle

%%%%%%%%%%%%%%%%%%%%%%%%%%%%%%%%%
%--------Introduction------------------
%%%%%%%%%%%%%%%%%%%%%%%%%%%%%%%%%
\noindent \textit{I. Introduction -} The magneto-optic Kerr effect (MOKE) \cite{JohnKerrLL.D..1877}, being magnetic circular dichroism and birefringence upon reflection from a magnetized sample, serves as a powerful tool in the area of thin-film magnetic characterization \cite{E.R.Moog.1985, Qiu.2000}. So far, applications such as vectorial magnetometry \cite{Florczak.1990, Vavassori.2000}, MOKE spectroscopy \cite{Harp.1993, Veis.2014} and microscopy \cite{Schafer.2007, JeffreyMcCord.2015} as well as time-resolved MOKE \cite{Beaurepaire.1996, Kirilyuk.2010} mainly rely on linear MOKE (LinMOKE), being linearly proportional to the magnetization $\bm{M}$. \\
In addition, quadratic-in-magnetization MOKE (QMOKE) \cite{Metzger.1965, JHamrle.2007}, being magnetic linear dichroism and birefringence upon reflection proportional to $\bm{M}^2$, can be employed to study antiferromagnets, for example in time-resolved experiments \cite{Saidl.2017, Zheng.2018, Zhao.2021}. QMOKE showed to be sensitive to the structural ordering of Heusler compounds \cite{Wolf.2011, Silber.2020} and to spin-orbit torque effects \cite{Montazeri.2015}. Vectorial magnetometry including QMOKE \cite{Mewes.2004, TKuschel.2011, Tesarova.2012, Tesarova.2013, Jang.2020}, QMOKE spectroscopy \cite{Silber.2020, Sepulveda.2003, Hamrlova.2016, RobinSilber.2018, Silber.2019}, as well as QMOKE microscopy \cite{Janda.2018, Xu.2019} have been realized. \\
However, any higher-order MOKE, such as cubic-in-magnetization MOKE (CMOKE), being proportional to $\bm{M}^3$, has only been mentioned very rarely so far \cite{Gridnev.1997, Petukhov.1998, K.Postava.2004} and has not yet been systematically investigated in a combined theoretic and experimental study. Furthermore, higher-order transport effects have been detected in magnetoresistive and anomalous Hall effects \cite{Limmer.2006, Limmer.2008, Meyer.2015} and can be expected analogously for magnetothermal and anomalous Nernst effects. Those effects possess equal symmetries as in the case of MOKE. However, MOKE offers advantages for spectroscopy, microscopy and time-resolved experiments that electrical measurements cannot provide such as spectral range, spatial and time resolution. \\
In this letter, we report on the observation of third-order-in-magnetization MOKE in Ni(111) thin films within a combined systematic theoretic and experimental study. We identify this CMOKE as a threefold in-plane angular dependence of the magnetically saturated longitudinal MOKE (LMOKE) response which cannot be attributed to LinMOKE or QMOKE, solely. For the quantitative description of our experimental data, we expand the Taylor series of the permittivity tensor up to third order in~$\bm{M}$. We analyze the impact of structural domain twinning and find that the threefold angular dependence of CMOKE is linearly decreasing with the degree of twinning. Thus, the observed MOKE contributions will be important for future applications related to structural domain twinning analysis, such as the investigation of huge magnetostriction in Ni$_2$MnGa \cite{Musiienko.2021}. \\

%%%%%%%%%%%%%%%%%%%%%%%%%%%%%%%%%
%--------THEORY------------------
%%%%%%%%%%%%%%%%%%%%%%%%%%%%%%%%%
\noindent \textit{II. Theoretic description -} MOKE can be related to the magnetically perturbed permittivity of the ferromagnetic layer. Hence, the Kerr angles $\Phi_{s/p}$ for $s$- and $p$-polarized incident light write \cite{Hamrle.2003} 
\begin{align}
\label{Kerr_analyt_s}%
\Phi_s &{}=\theta_{s}+i\epsilon_{s}= A_{s}\left(\varepsilon_{yx}-\frac{\varepsilon_{yz}\varepsilon_{zx}}{\varepsilon_{d}}\right)+B_s\varepsilon_{zx} \text{  ,} \\
\Phi_p &{}=\theta_{p}+i\epsilon_{p}= -A_{p}\left(\varepsilon_{xy}-\frac{\varepsilon_{xz}\varepsilon_{zy}}{\varepsilon_{d}}\right)+B_p\varepsilon_{xz} \text{  ,}
\label{Kerr_analyt_p}%
\end{align}
in which $\theta_{s/p}$ is the Kerr rotation and $\epsilon _{s/p}$ is the  Kerr ellipticity.  $A_{s/p}$ and $B_{s/p}$ are the optical weighting factors which are even and odd functions of the angle of incidence, respectively. $\varepsilon _d$ represents the non-magnetic diagonal permittivity tensor elements of a cubic crystal structure. \\
In order to describe QMOKE, the permittivity tensor $\bm{\varepsilon }$ is typically developed up to second order in~$\bm{M}$ \cite{Visnovsky.2006, JHamrle.2007, Sepulveda.2003}. However, to successfully describe all experimental observations presented in our study, we have to develop $\bm{\varepsilon }$ up to third order in~$\bm{M}$. Using the Einstein summation, the elements of $\bm{\varepsilon }$ are expressed as
\newpage
\begin{align}
	\varepsilon_{ij}&=\varepsilon_{ij}^{(0)}  +  K_{ijk}M_k  +  G_{ijkl} M_k M_l + H_{ijklm} M_k M_l M_m \text{ .} 
\label{Perm_2nd_order}
\end{align}
Here, $\varepsilon_{ij}^{(0)}$ describes the elements of the non-perturbed permittivity tensor. $M_{k/l/m}$ are components of the normalized magnetization~$\bm{M}$, and $K_{ijk}$ and $G_{ijkl}$ are the components of the linear and quadratic magneto-optic (MO) tensors $\bm{K}$ and $\bm{G}$ \cite{Visnovsky1986}, respectively. We further add the components $H_{ijklm}$ of the cubic MO five-rank tensor $\bm{H}$ \cite{K.Postava.2004}. \\
Using the Onsager relation $\varepsilon _{ij}(\bm{M})=\varepsilon _{ij}(-\bm{M})$ as well as symmetry arguments of cubic crystal structures, these tensors can be simplified in such a way that $\bm{K}$ is solely determined by one independent parameter $K$, while $\bm{G}$ is described by two independent parameters, namely $G_s=(G_{11}-G_{12})$ and $2G_{44}$ \cite{Visnovsky.2006, Silber.2019}. The difference of these two parameters $\Delta G = G_s-2G_{44}$ denotes the anisotropic strength of $\bm{G}$ \cite{JHamrle.2007, TKuschel.2011}. $\bm{H}$ possesses two independent parameters \cite{Petukhov.1998} which can be named $H_{123}$ and $H_{125}$ following the derivation in Ref. \cite{Petukhov.1998}. Moreover, we introduce an anisotropy parameter $\Delta H = H_{123}-3H_{125}$ which describes the anisotropic strength of $\bm{H}$. \\
In order to yield the perturbed permittivity tensor for a (111)-oriented cubic crystal structure, the permittivity tensor is rotated around the $\hat{x}$-axis of our coordinate system (see Supp. Mat. \cite{Supp}) by $\vartheta _{[100]} = 45^{\circ }$ and subsequently around the $\hat{y}$-axis by $\vartheta _{[010]} = \arcsin(1/\sqrt{3}) \approx  35.26^{\circ }$. To describe the in-plane sample rotation in our experiment, we further apply a rotation to the permittivity tensor by the general angle $\alpha $ around the $\hat{z}$-axis of our coordinate system. The respective expressions for the permittivity tensor elements are written down in the Supplemental Materials \cite{Supp}. \\

%%%%%%%%%%%%%%%%%%%%%%%%%%%%%%%%%
%--------Eight-directional method------------------
%%%%%%%%%%%%%%%%%%%%%%%%%%%%%%%%%
\noindent \textit{III. Eight-directional method -} In order to analyze individual MOKE contributions with different dependencies on $\bm{M}$ experimentally, a separation algorithm, known as the eight-directional-method \cite{Postava.2002}, can be used. So far, this method or similar ones have been utilized to characterize (001)- and (011)-oriented thin films of cubic crystal structure \cite{Silber.2019, Liang.2016} while the QMOKE in (111)-oriented thin films so far has been identified only by magnetization loop symmetrization \cite{PKMuduli.2009} and theoretically \cite{Hamrlova.2013}. During the eight-directional method, the MOKE signal is measured for eight different in-plane magnetization directions $\mu = 0^{\circ } + k\cdot 45^{\circ }, k = \{0,1,...,7\}$ (see Supp. Mat. \cite{Supp}). From these measurements, four different MOKE contributions are extracted. Analytical equations for each of the four MOKE contributions can be derived by inserting the permittivity tensor elements into the Eqs. (\ref{Kerr_analyt_s}) and (\ref{Kerr_analyt_p}). Note that we only consider contributions up to third order in $\bm{M}$ and stick to in-plane magnetization. \\
The first MOKE contribution, separated by the eight-directional method, is the LinMOKE contribution \linebreak ($\propto M_L$) together with the longitudinal CMOKE contribution ($\propto M_L^3$) for the magnetically saturated longitudinal magnetization component $M_L$ (in-plane and parallel to the plane of incidence)
\begin{align}
    \Phi _{M_L, M_L^3} &= \frac{1}{2}\left( \Phi _{s/p}^{\mu = 90^\circ } - \Phi _{s/p}^{\mu = 270^\circ } \right) \nonumber \\
    & = \pm B_{s/p} \left(K + \frac{H_{123}+3H_{125}}{2} \right) \nonumber \\
    & \quad -A_{s/p} \frac{\sqrt{2}}{6}\left( \Delta H + \frac{K\Delta G}{\varepsilon _d } \right) \sin{(3\alpha)} \text{  .}
    \label{ML}
\end{align}
Here, we can identify an isotropic term $\propto $~$M_L$ and $\propto $~$M_L^3$ as well as an anisotropic threefold term $\propto $~$M_L^3$. The amplitude of the anisotropic contribution is stemming from two MO parameters, being $\Delta H$ and the product $K \Delta G$. The latter stems from the intermixing of $\varepsilon_{yz}\varepsilon_{zx}$ and $\varepsilon_{xz}\varepsilon_{zy}$ in Eqs. (\ref{Kerr_analyt_s}) and (\ref{Kerr_analyt_p}). \\
The second MOKE contribution is the transverse CMOKE contribution ($\propto M_T^3$) of the magnetically saturated transverse magnetization component $M_T$ (in-plane and perpendicular to the plane of incidence)
\begin{align}
    \Phi _{M_T^3} &= \frac{1}{2}\left( \Phi _{s/p}^{\mu = 0^\circ } - \Phi _{s/p}^{\mu = 180^\circ } \right) \nonumber \\
    & = A_{s/p} \frac{\sqrt{2}}{6}\left( \Delta H + \frac{K\Delta G }{\varepsilon _d } \right) \cos{(3\alpha)}
    \label{MT}
\end{align}
which only consists of an anisotropic threefold term $\propto $~$M_T^3$. In contrast to $\Phi _{M_L, M_L^3}$, $\Phi _{M_T^3}$ depends on $\cos{(3\alpha )}$ instead of $\sin{(3\alpha )}$, but has the same amplitude. \\ The third MOKE contribution is the QMOKE contribution ($\propto M_LM_T$)
\begin{align}
    \Phi _{M_LM_T} &= \frac{1}{2}\left( \Phi _{s/p}^{\mu = 45^\circ } + \Phi _{s/p}^{\mu = 225^\circ } - \Phi _{s/p}^{\mu = 135^\circ } - \Phi _{s/p}^{\mu = 315^\circ } \right) \nonumber \\
    &= \pm A_{s/p}\left( 2G_{44} + \frac{1}{3}\Delta G - \frac{K^2}{\varepsilon _d}\right) \nonumber \\
    &\quad -B_{s/p}\frac{\sqrt{2}}{3} \Delta G \sin{(3\alpha )}
    \label{MLMT}
\end{align}
which consists of an anisotropic threefold term and an isotropic term that depends on the MO parameters of $\bm{G}$ as well as $K^2$. \\
The fourth and final MOKE contribution, separated by the eight-directional method, is the QMOKE contribution ($\propto M_T^2-M_L^2$)
\begin{align}
    \Phi _{M_T^2-M_L^2} &= \frac{1}{2}\left( \Phi _{s/p}^{\mu = 0^\circ } + \Phi _{s/p}^{\mu = 180^\circ } - \Phi _{s/p}^{\mu = 90^\circ } - \Phi _{s/p}^{\mu = 270^\circ } \right) \nonumber \\
    &= -B_{s/p}\frac{\sqrt{2}}{3}\Delta G \cos{(3\alpha )}
    \label{MT2ML2}
\end{align}
which only consists of an anisotropic threefold term. While $\Phi _{M_T^2-M_L^2}$ depends on $\cos{(3\alpha )}$, the anisotropic part of $\Phi _{M_LM_T}$ depends on $\sin{(3\alpha )}$, but with the same amplitude determined by~$\Delta G$.
\begin{figure*}
\begin{center}
\includegraphics{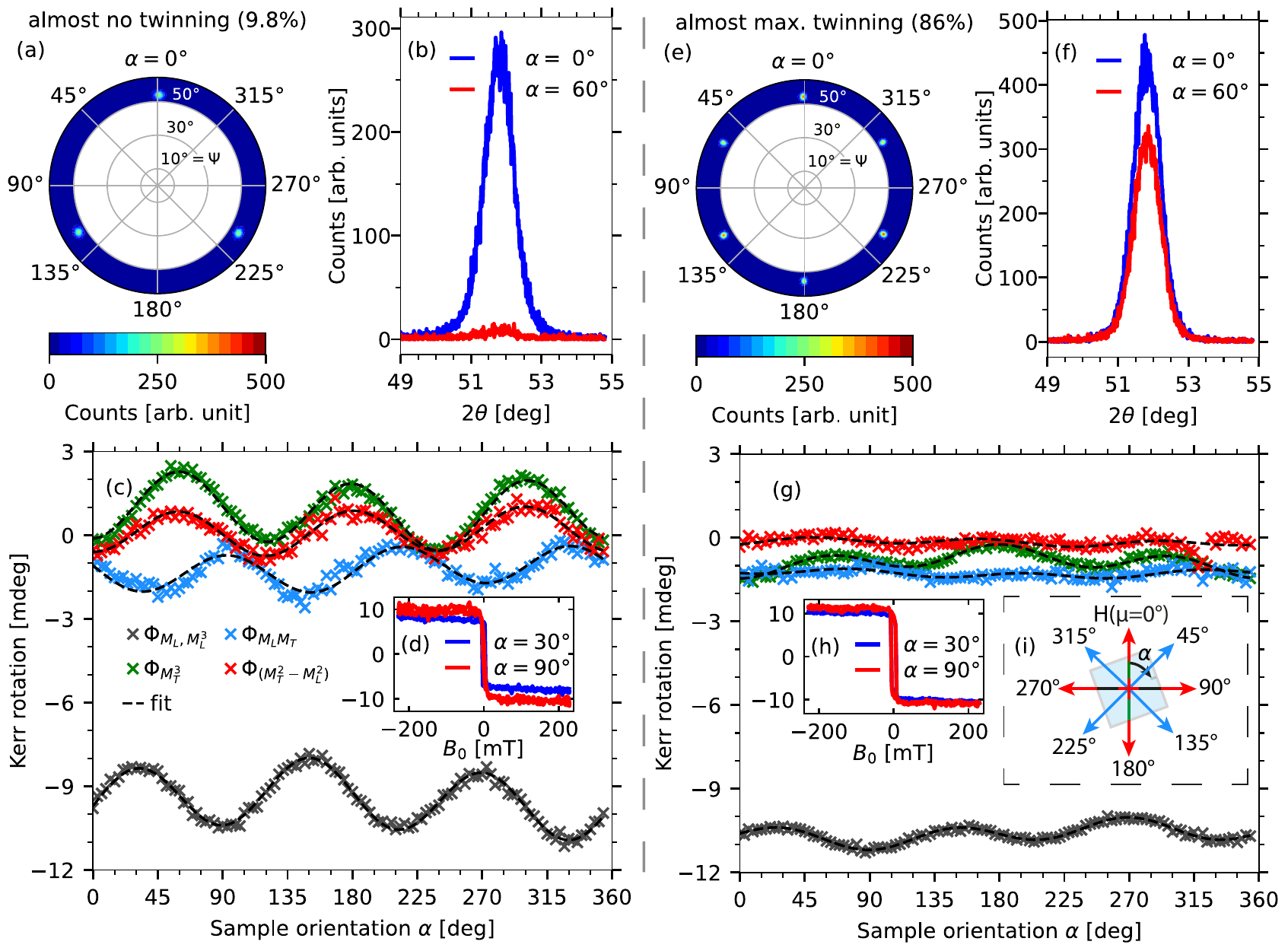}
\end{center}
\vspace{-7mm}
\caption{XRD and MOKE data of the Ni(111) samples with almost no (9.8\%) twinning (left) and almost maximum (86\%) twinning (right). (a,e) Off-specular XRD scan (Euler's cradle texture map) showing the Ni\{200\} peaks at $2\theta = 51.832^\circ $. (b,f)~XRD $\theta - 2\theta $ scans of the Ni\{200\} peaks at selected $\alpha $ angles. (c,g) Eight-directional method performed with a magnetic field strength of 230\,mT using a wavelength of 635\,nm. The dependence of four MOKE contributions on the sample orientation $\alpha $ is demonstrated. (d,h) Magnetization curves for exemplary sample orientations. (i) Schematic displaying the eight directions in which the magnetic field is applied to measure the different MOKE contributions.}
\vspace{-3mm}
\label{Ni111}
\end{figure*} \\

%%%%%%%%%%%%%%%%%%%%%%%%%%%%%%%%%
%--------Experimental background------------------
%%%%%%%%%%%%%%%%%%%%%%%%%%%%%%%%%
\noindent \textit{IV. Experimental background -} The fcc Ni(111) thin films were prepared by magnetron sputtering onto MgO(111) substrates in an Ar atmosphere of $2.1 \times 10^{-3}$ mbar. Parts of the growth procedure have been adapted from Ref. \cite{PerSandstrom.1999}. The substrates have been ultrasonically cleaned in acetone, isopropanol and ethanol and then dried in $N_2$. Afterwards, they were annealed in vacuum at 800$^\circ $C. The cleaning and annealing times were varied throughout the samples series to achieve different degrees of twinning. The substrates of the exemplarily displayed samples 1 and 2 were annealed for 1h and 15 minutes, respectively (see Supp. Mat. \cite{Supp} for more details). About 20\,nm thick Ni layers were then deposited onto the substrates with a growth rate of 0.066\,nm/s at a substrate temperature of 350$^\circ $C. Afterwards, the Ni layers were capped with approximately 3\,nm of Si which was grown at room temperature. \\
The Ni layer thicknesses were confirmed by X-ray reflectivity measurements (see Supp. Mat. \cite{Supp}), using the Cu $K_{\alpha }$ source of a Phillips X'pert Pro MPD PW3040-60 diffractometer. Specular X-ray diffraction (XRD) $\theta $-$2\theta $ scans (see Supp. Mat. \cite{Supp}) as well as off-specular XRD texture mapping using an Euler cradle were employed to examine the crystalline growth of the samples. The texture measurements were performed for 360$^\circ $ of sample rotation $\alpha $ with a tilt of the samples $\Psi =\langle 50 ^\circ $, $60^\circ \rangle $, effectively capturing the Ni\{200\} diffraction peaks. Since the Ni layer is grown in a (111) orientation, there are two growth types possible (structural domain twinning) which differ from each other by an in-plane rotation of 60$^\circ $. Each (111) phase produces a diffraction pattern with threefold symmetry, but with a $\Delta \alpha =60^\circ $ difference between the patterns. \\
MOKE measurements were carried out using \textit{p}-polarized light of 635\,nm wavelength and a 45$^{\circ }$ angle of incidence. Both the samples as well as the magnetic field were rotated in the sample plane. More detailed descriptions of the MOKE setup can be found in Refs. \cite{Kehlberger.2015, Muglich.2016}. \\

%%%%%%%%%%%%%%%%%%%%%%%%%%%%%%%%%
%--------Results------------------
%%%%%%%%%%%%%%%%%%%%%%%%%%%%%%%%%
\noindent \textit{V. Results -} 
The off-specular XRD texture mapping of sample 1 is presented in Fig. \ref{Ni111}(a). The scan was performed at 2$\theta $=51.832$^\circ $ and thus shows the Ni\{200\} peaks. Additional $\theta $-2$\theta $ scans at all Ni\{200\} peak positions were made and are displayed exemplarily in Fig. \ref{Ni111}(b). The appearance of only three intense Ni\{200\} peaks indicates that almost no twinning occurs in sample~1. The integrated peak intensities of the dominant phase make up 95.1\% of the combined peak intensities which lead to a degree of twinning of $100\%-(95.1\%-4.9\%)=9.8\%$. \\
The in-plane angular dependencies of the four MOKE contributions are measured by the eight-directional method according to Eqs. \eqref{ML}-\eqref{MT2ML2} and presented in Fig. \ref{Ni111}(c). The experimentally observed dependencies are in good agreement with the analytical equations. In all four MOKE contributions, pronounced threefold angular dependencies are visible. While $\Phi_{M_L, M_L^3}$ and $\Phi_{M_LM_T}$ follow a $\sin{(3\alpha )}$ curve (see Eqs. \eqref{ML} and \eqref{MLMT}), $\Phi_{M_T^3}$ and $\Phi_{M_T^2-M_L^2}$ describe a $\cos{(3\alpha )}$ behaviour (see Eqs. \eqref{MT} and \eqref{MT2ML2}). In addition, the expected offsets of $\Phi_{M_L, M_L^3}$ and $\Phi_{M_LM_T}$ can be identified (see Eqs. \eqref{ML} and \eqref{MLMT}), while $\Phi_{M_T^2-M_L^2}$ has a vanishing offset (see Eq. \eqref{MT2ML2}). $\Phi_{M_T^3}$ shows a finite offset which is not predicted by theory (see Eq. \eqref{MT}). Magnetic field misalignments as a possible origin of this offset have been excluded (see Supp. Mat. \cite{Supp}). \\
A small superimposed onefold angular dependence can be seen mainly in the $\Phi_{M_L, M_L^3}$ and $\Phi_{M_T^3}$ contributions. We attribute it to a vicinal MOKE (VISMOKE) contribution \cite{Hamrle.2003} caused by a slight miscut of the substrate (see Supp. Mat. \cite{Supp} for further details). Note that the sample possesses a uniaxial magnetic anisotropy with the magnetic easy axis oriented parallel to the step edges of the vicinal surface (see Supp. Mat. \cite{Supp}). Therefore, it can be concluded that the three-fold anisotropies in $\Phi_{M_L, M_L^3}$ and $\Phi_{M_T^3}$ do not stem from the uniaxial magnetic anisotropy of the sample. \\
The off-specular XRD texture mapping and $\theta $-2$\theta $ scans of the Ni\{200\} peaks of sample 2 are displayed in Figs. \ref{Ni111}(e,f). Here, six peaks of similar intensity can be identified. The dominant phase makes up 57.0\% of the combined peak intensities related to 86\% of twinning. In all four MOKE contributions displayed in Fig. \ref{Ni111}(g), the threefold angular dependencies have undergone a clear reduction, compared to sample~1. This reduction of the Q- and CMOKE amplitudes is due to the fact that the threefold angular dependencies of both (111) phases cancel each other out due to their $\Delta \alpha =60^\circ $ difference in orientation.  Exemplary magnetization curves at $\alpha = 30^\circ ,90^\circ $ are displayed for both samples in Figs. \ref{Ni111}(d) and \ref{Ni111}(h), respectively. A stronger anisotropy in the saturation values can be identified in sample 1 compared to sample 2 due to the lower degree of twinning. \\
\begin{figure}
\begin{center}
\includegraphics{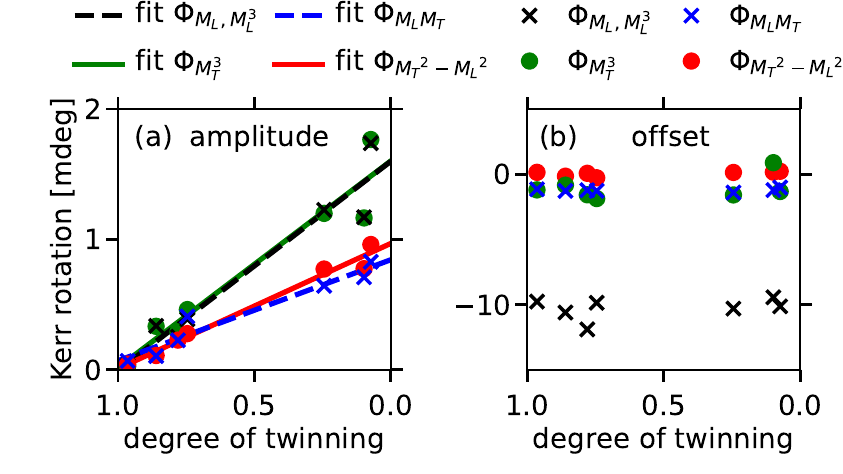}
\end{center}
\vspace{-7mm}
\caption{Dependence of (a) amplitudes and (b) offsets of the individual MOKE contributions on the degree of structural domain twinning. The dependence of the amplitudes on the degree of twinning is fitted using a linear regression. }
\vspace{-3mm}
\label{twinning_dependencies}
\end{figure}

\noindent In order to analyze offsets, amplitudes and phase dependencies quantitatively, fits of the form $C + A_1 \sin{(3\alpha +\alpha _1)} + A_2 \sin{(\alpha + \alpha _2)}$ have been performed (see dashed lines in Figs. \ref{Ni111}(c) and \ref{Ni111}(g)). Here, the onefold sine contribution is related to VISMOKE and the phase shift $\alpha _2$ correlates with the miscut direction of the respective Ni film (see Supp. Mat. \cite{Supp}). The phase shifts $\alpha _1$ of all contributions are in good agreement to each other with additional 90$^\circ $ offset for $\alpha _1$ of $\Phi_{M_T^3}$ and $\Phi_{M_T^2-M_L^2}$ due to the $\cos{(3\alpha )}$ dependence instead of $\sin{(3\alpha )}$ (see Eqs. \eqref{ML}-\eqref{MT2ML2}). \\
As predicted by theory, both CMOKE contributions $\Phi _{M_L,M_L^3}$ and $\Phi _{M_T^3}$ have similar amplitudes of the threefold angular dependencies. For sample 1, these are $(1.17 \pm 0.01)$\,mdeg and $(1.16 \pm 0.02)$\,mdeg. In sample 2, both of these amplitudes are reduced to $(0.33 \pm 0.01)$\,mdeg. The QMOKE contributions $\Phi _{M_LM_T}$ and $\Phi _{M_T^2-M_L^2}$ also show similar amplitudes for each sample. In sample 1, these are $(0.71 \pm 0.03)$\,mdeg and $(0.77 \pm 0.02)$\,mdeg. In sample 2, these amplitudes are reduced to $(0.10 \pm 0.01)$\,mdeg and $(0.11 \pm 0.02)$\,mdeg. \\
An overview of the amplitudes and offsets in all samples is displayed in Fig. \ref{twinning_dependencies}. It is clearly visible that the amplitudes of the angular dependencies stemming from CMOKE and QMOKE are directly proportional to the amount of structural domain twinning in the sample. The anisotropic MOKE contributions vanish in a fully twinned sample and are maximal in a single crystalline sample. The offsets of $\Phi_{M_L, M_L^3}$ and $\Phi _{M_LM_T}$ are independent from the degree of twinning. Slight variations between the offsets can be attributed to minor variations in layer thicknesses. \\
Measurements of the Kerr ellipticity and with $\lambda = 406$\,nm have also been conducted for two samples using the eight-directional method (see Supp. Mat. \cite{Supp}). The same reduction of CMOKE and QMOKE amplitudes with the degree of twinning has been observed, while the offsets stay constant. In the Supp. Mat. \cite{Supp} the dependencies of the MOKE contributions on the magnetic field strength are also shown. While the offsets of all contributions are constant once the sample is magnetically saturated, the CMOKE amplitudes slightly decrease with increasing magnetic field. \\

\noindent \textit{VI. Numerical simulations -} The CMOKE threefold angular dependence can be described qualitatively just by $K$\,$\Delta G$ (see Eqs. \eqref{ML} and \eqref{MT}). In order to check whether the $\bm{H}$ tensor is needed to quantitatively describe the amplitude of the CMOKE oscillations, numerical simulations using Yeh's transfer matrix formalism \cite{PochiYeh.1980} with and without the tensor $\bm{H}$ have been conducted. In the numerical model, the MO parameters are set as free parameters and the model is fitted to the experimental data.
\begin{figure}[]
\begin{center}
\includegraphics{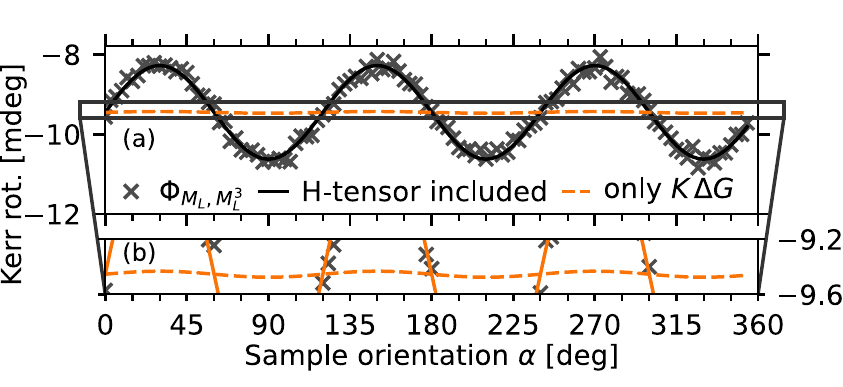}
\end{center}
\vspace{-7mm}
\caption{$\Phi _{M_L, M_L^3}$ contribution of sample 1 (VISMOKE subtracted). (a) Simulations with and without the permittivity tensor of third order in $\bm{M}$ are shown. (b) Close-up of the simulation without the $\bm{H}$ tensor.}
\vspace{-3mm}
\label{ML_simulation}
\end{figure} \\
In Fig. \ref{ML_simulation}(a), simulations with and without $\bm{H}$ tensor are shown for sample 1. Although the fit was performed for all contributions of the eight-directional method simultaneously, we only show the results for $\Phi _{M_L,M_L^3}$ here (see Supp. Mat. \cite{Supp} for the numerical simulation of all MOKE contributions). The superimposed onefold angular dependency due to VISMOKE has been subtracted prior to the numerical fitting. The simulation of MOKE up to second order in $\bm{M}$ (i.e. without $\bm{H}$) shows negligible angular dependencies of $\Phi _{M_L,M_L^3}$ (see Fig. \ref{ML_simulation}(b)). In order to describe $\Phi _{M_L,M_L^3}$ solely by $K$ and $\Delta G$ in the numerical model, the calculated amplitudes for $\Phi _{M_LM_T}$ and $\Phi _{M_T^2-M_L^2}$ would be largely above the experimental values. However, with the $\bm{H}$ tensor included in the simulation, CMOKE and the QMOKE contributions can be fitted well, simultaneously.
Therefore, the  contribution which is stemming from $K\,\Delta G$ can be assumed to be small compared to the contribution stemming from $\Delta H$. This shows that the $\bm{H}$ tensor is indispensable for a quantitative description of CMOKE in Ni(111) films. \\

%%%%%%%%%%%%%%%%%%%%%%%%%%%%%%%%%
%--------Conclusion------------------
%%%%%%%%%%%%%%%%%%%%%%%%%%%%%%%%%
\noindent \textit{VII. Conclusion -} We observed MOKE contributions of third order in $\bm{M}$, called CMOKE, in Ni(111) thin films. These contributions manifest themselves in strong threefold angular dependencies of e.g. LMOKE which are even larger than the ones of the QMOKE contributions. We were able to describe CMOKE quantitatively by describing the permittivity tensor up to third order in $\bm{M}$ and including the fifth-rank tensor $\bm{H}$. Furthermore, we showed that the angular dependencies of the different MOKE contributions are suppressed in case of structural domain twinning. Our findings can be used in new applications (spectroscopy, microscopy, pump-probe) based on CMOKE, e.g. for the analysis of structural domain twinning.

\vspace{-5mm}
\section*{Acknowledgments}
\vspace{-5mm}
\noindent This article has been produced with the financial support of the European Union under the REFRESH - Research Excellence For Region Sustainability and High-tech Industries project number CZ.10.03.01/00/22$\_$003/0000048 via the Operational Programme Just Transition. Additionally, the infrastructure used was made available through project No. CZ02.01.01/00/22$\_$008/0004631- "Materials and Technologies for Sustainable Development," funded by the European Union and the state budget of the Czech Republic within the framework of the Jan Amos Komensky Operational Program. \\
J.H. acknowledges support by the European Union project Matfun, Project No. CZ.02.1.01/0.0/0.0/15$\_$003/0000487.  M.G. acknowledges support from the BMBF project DiProMag (Plattform MaterialDigital, Project No. 13XP5 120B).

\bibliography{main.bib}

\end{document}